\documentclass[prd,onecolumn,groupedaddress,nofootinbib,english,floatfix,superscriptaddress,preprintnumbers,notitlepage]{revtex4-1}
\usepackage{graphicx}
\usepackage{cancel}
\usepackage{amssymb}
\usepackage{textcomp}
\usepackage{amsmath}
\usepackage{mathtools}
\usepackage{bm}
\usepackage{times}
\usepackage{epsfig}
\usepackage[dvipsnames]{xcolor}
\usepackage{graphics}
\usepackage{hyperref}
\usepackage{setspace}
\usepackage{comment}
\usepackage{subcaption}
\usepackage[utf8]{inputenc}
\usepackage[english]{babel}
\usepackage{textcomp}
\usepackage{braket}
\usepackage[section]{placeins}
\usepackage{siunitx}
\usepackage{latexsym}
\usepackage{mathrsfs}

\usepackage{tikz}
\usetikzlibrary{patterns}
\usetikzlibrary{shapes.geometric}
\usetikzlibrary{decorations.pathmorphing}
\usetikzlibrary{arrows.meta}
\tikzset{snake it/.style={decorate, decoration=snake}}
\usetikzlibrary{positioning}
\usetikzlibrary{arrows,shapes,positioning}
\usetikzlibrary{decorations.markings}
\tikzstyle arrowstyle=[scale=1]
\tikzstyle directed=[postaction={decorate,decoration={markings,mark=at position .65 with {\arrow[arrowstyle]{stealth}}}}]
\tikzstyle reverse directed=[postaction={decorate,decoration={markings,mark=at position .65 with {\arrowreversed[arrowstyle]{stealth};}}}]

\tikzset{->-/.style={decoration={
  markings,
  mark=at position #1 with {\arrow{>}}},postaction={decorate}}}

\tikzset{-<-/.style={decoration={
  markings,
  mark=at position #1 with {\arrow{<}}},postaction={decorate}}}

\newcommand{\rU}{{\mathrm{U}}}
\newcommand{\rI}{{\mathrm{I}}}
\newcommand{\rII}{{\mathrm{II}}}
\newcommand{\rIII}{{\mathrm{III}}}
\newcommand{\rIV}{{\mathrm{IV}}}

\newcommand{\mH}{\ensuremath{\mathcal{H}}}
\newcommand{\mHc}{\ensuremath{\mathcal{H}_c}}
\newcommand{\mCH}{\ensuremath{\mathcal{CH}}}
\newcommand{\ome}{\ensuremath{\omega_{\rI}}}

\newcommand{\ri}{\ensuremath{\mathrm{in}}}
\newcommand{\ru}{\ensuremath{\mathrm{up}}}

\newcommand{\del}{{\partial}}

\DeclareMathOperator{\Imag}{Im}
\DeclareMathOperator{\Real}{Re}

\begin{document}
\title{Renormalized charged scalar current in the Reissner-Nordström-de Sitter spacetime}

\author{Christiane Klein}
\email{klein@itp.uni-leipzig.de}
\author{Jochen Zahn}
\email{jochen.zahn@itp.uni-leipzig.de}
\affiliation{Institut f\"ur Theoretische Physik, Universit\"at Leipzig,\\ Br\"uderstra{\ss}e 16, 04103 Leipzig, Germany}

\begin{abstract}
We perform a Hadamard point-split renormalization of the current density $j_\nu$ of a charged scalar field in a Reissner-Nordstr{\"om}-de Sitter spacetime. We compute numerically the expectation values of the components $j_r$ and $j_t$ in the Unruh state in the exterior region and study their dependence on the field parameters and the position. 
%In this work, we calculate the current of a massive charged quantum scalar field in a Reissner-Nordström-de Sitter spacetime, applying angular point-splitting and Hadamard regularization. We find that the counterterm is finite and vanishes on the horizons. The expression for the current is evaluated numerically. Its functional dependence on the mass and charge of the scalar field is in agreement with estimates from the Schwinger formula applied previously to black hole discharge.
\end{abstract}
\maketitle
\section{Introduction}
In astrophysical contexts, it is expected that any charged black hole (BH) will quickly lose its charge by accreting matter of the opposite charge. But even an isolated charged BH loses its charge by quantum effects. 
The discharge of charged BHs by Hawking radiation or pair creation due to the electromagnetic background field has already been anticipated by Hawking himself in his seminal work \cite{Hawking:1974}. This effect has been studied further, for example, by \cite{ Zaumen:1974,Carter:1974,Damour:1974, Gibbons:1975,Hiscock:1990,Herman:1994,Herman:1995, Khriplovich:1999,Gabriel:2000,Chen:2012,Kim:2013,Johnson:2019,Ong:2019,Ong:2019a, Balakumar:2020}. These works typically focus on the discharge in the exterior region and, as far as quantitative results are concerned, on large field masses and charges. While this is certainly the most physically interesting phenomenon and parameter range, we think that there are still interesting open questions. Here, we want to answer a few of them.

The study of discharge in the exterior region typically proceeds either by considering pair creation (for example using Schwinger's pair creation rate \cite{Schwinger:1951}), or by computing the expectation value of the current $j_r$. By charge conservation, this quantity is determined by its value at an arbitrary position in stationary and rotationally symmetric states. We compute it numerically at the event horizon in the Unruh state, albeit only for small mass $\mu$ and charge $q$ of the field and a large cosmological constant $\Lambda$. 
We find an exponential decay in the mass and a proportionality $\sim q^2$ for small $q$.
%We verify the exponential decay in the mass expected from Schwinger's pair creation formula and find a proportionality $\sim q^2$ for small $q$. 

Another potentially interesting observable, which to the best of our knowledge has not yet been considered in the literature, is the current component $j_t$, representing, up to a negative factor, the (vacuum polarization) charge density, as measured by a stationary observer. Certainly, for astrophysical BHs and realistic field masses, one expects this polarization charge density to decay rapidly, as for QED in an external Coulomb potential (exponentially according to Uehling's result linear in the external potential \cite{Uehling:1935uj} or subexponentially when treating the external potential nonperturbatively \cite{Mohr:1998grz}). It would be nice to verify this intuition. We make first steps in this direction, as we calculate a finite counterterm from a Hadamard point-split renormalization. We also compute the vacuum polarization numerically in the Unruh state, again in the regime of small mass $\mu$ and charge $q$ of the field and a large cosmological constant $\Lambda$.

The inclusion of a cosmological constant, i.e., working on a Reissner-Nordstr{\"o}m-de Sitter (RNdS) instead of a Reissner-Nordstr{\"o}m (RN) spacetime, goes beyond previous work (focusing on the asymptotically flat case) and has further advantages: numerical calculations are easier, as the corresponding mode equation has only regular singular points (instead of an irregular singular point at $\infty$ for RN). This allows for an efficient numerical scheme \cite{Hollands:2020}, which works best if the cosmological and event horizons are of the same order of magnitude (corresponding to a large cosmological constant). Furthermore, (quantum) fields on RNdS have attracted considerable attention in recent years, due to the possibility of violations of strong cosmic censorship (sCC) \cite{Penrose:1974} on such spacetimes \cite{Cardoso:2017, Dias:2018, Cardoso:2018}.

Our results are also relevant regarding the ongoing debate about sCC. The sCC conjecture roughly states that at the inner horizon of a charged or rotating black hole, like $\mCH^R$ in Fig. \ref{fig:RNdS}, local observables such as the current of a charged scalar field will diverge. This prevents any observer from crossing the inner horizon and entering the region beyond it, where determinism breaks down. It was suspected long ago \cite{BirrellDavies:1978} and recently shown \cite{Hollands:2019} that the stress tensor of a real scalar field diverges in a universal way at the inner horizon. The divergence is stronger than that of classical fields and strong enough to ensure the validity of sCC.

In the context of charged BHs, the consideration of charged fields is natural. In view of the sCC conjecture, it is then interesting to study the behavior of the expectation value of the current density $j_\nu$ near the inner horizon. Our results on the point-split renormalization of the current, which are also valid in the interior region, are used in \cite{PRLpaper} to do this.

The study of charged scalars on RNdS is made slightly more complicated by the fact that an instability is present in the regime of small mass $\mu$ and charge $q$ of the field and small cosmological constant $\Lambda$ \cite{Zhu:2014,Konoplya:2014,Dias:2018, Hod:2018}. As far as our general results on the point-split renormalization of the current density in the Unruh state are concerned, we assume that we are in a parameter range in which the instability is absent. For our concrete computations, we choose parameters such that the instability is avoided (this is one of the reasons for choosing a large cosmological constant $\Lambda$).

Let us comment on the relation to previous similar work. In \cite{Herman:1995}, an expression for the Hadamard point-split renormalization of the current in general spacetimes and external background on-shell electromagnetic fields is derived (though not evaluated). A divergent counterterm from Hadamard point splitting is obtained, in contrast to our result of a finite counterterm. As explained below, this seems to be due to a sign mistake in \cite{Herman:1995}. Reference \cite{Balakumar:2019} has derived a prescription for the calculation of the renormalized current density of a charged scalar field in a curved spacetime with a background electromagnetic field via Hadamard point splitting, but only applies it to show that the renormalized current satisfies current conservation. Recently, the expectation value of $j_r$ was computed in the massless case for the Boulware state in a RN spacetime \cite{Balakumar:2020}. We go beyond these results in that we allow for a nonzero mass and cosmological constant, use the physically more relevant Unruh state, and also compute the expectation value of the component $j_t$, related to the charge density.

The rest of the paper is structured as follows. In Sec. \ref{sec:2}, we introduce the dynamics of the scalar field and the Unruh vacuum. These are used in Sec. \ref{sec:current} to derive a point-splitting regularized expression for the current in the Unruh state. The parametrix for the current is determined in Sec. \ref{sec:ct}. We present numerical results in Sec. \ref{sec:num} and conclude in Sec. \ref{sec:fin}.

Throughout, we work in units $\hbar=c=G=4\pi\epsilon_0=1$.

\section{Setup}
\label{sec:2}
We consider a charged scalar field $\Phi$ on a RNdS spacetime. 
\begin{figure}
%\begin{tikzpicture}[scale=1]
% \draw (-2,2) -- (-4,4) node[midway, below, sloped]{$\mathcal{H}^L$};
 %\draw (0,0) -- (2,2) node[midway, below, sloped]{$\mathcal{H}_c^-$};
 %\draw (2,2) -- (0,4) node[midway,above,sloped]{$\mHc^L$};
 %\draw (-2,2) -- (0,4) node[midway, above, sloped]{$\mathcal{H}^R$};
% \draw (0,4) -- (-2,6) node[midway, above, sloped]{$\mathcal{C}\mathcal{H}^R$};
% \draw (-4,4) -- (-2,6) node[midway, above, sloped]{$\mathcal{C}\mathcal{H}^L$};

% \draw[green, thick] (-4,4) .. controls (0,2) .. (4,4) node[midway,above]{$\Sigma$};

% \draw[snake it] (0,4) -- (0,8);
 %\draw (-2,6) -- (0,8);
% \draw[double=black] (0,4) -- (4,4); 
 %\draw (2,2) -- (4,4) node[midway, below, sloped]{$\mathcal{H}_c^R$};
% \draw (0,2) node{${\rm I}$};
% \draw (2,3) node{${\rm III}$};
 %\draw (-2,4) node{${\rm II}$};
 %\draw (-1,6) node{${\rm IV}$};
% \draw[densely dotted] (0,0) .. controls (-0.8,4) .. (0,5);

% \draw[fill=white] (-4,4) circle (2pt);
% \draw[fill=white] (4,4) circle (2pt);
 %\draw[fill=white] (0,8) circle (2pt);
% \draw[fill=black] (-2,6) circle (2pt);
% \draw[fill=white] (0,0) circle (2pt) node[below]{$i^-$};
% \draw[fill=black] (-2,2) circle (2pt);
 %\draw[fill=white] (0,4) circle (2pt);% node[above right]{$i^+$};
% \draw[fill=black] (2,2) circle (2pt);
 % \end{tikzpicture}
 \includegraphics[width=0.5\textwidth]{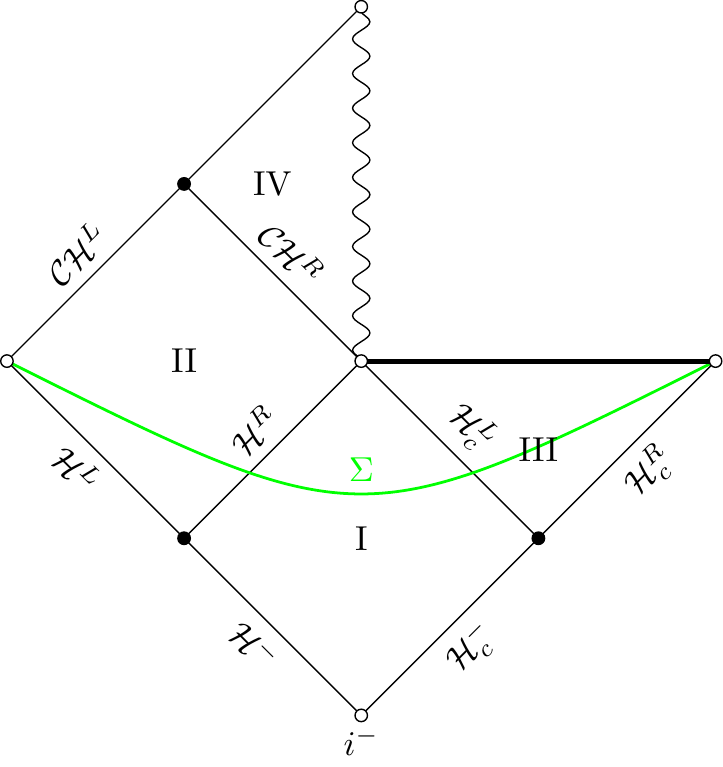}
\caption{Penrose diagram for  Reissner-Nordstr\"om-de Sitter spacetime. $\rI$ and  $\rIII$ constitute the exterior region, with $\rIII$ being the region beyond the cosmological horizon $\mHc$, i.e., out of causal contact with the BH interior. The latter is represented by regions $\rII$ and $\rIV$, which are separated from the exterior by the event horizon $\mH^R$ and from each other by the Cauchy horizon $\mCH^R$. The curvature singularity is represented by the wiggled line. The green line indicates a Cauchy surface $\Sigma$ for the region $\rI \cup \rII \cup \rIII$. Here we focus on region $\rI$ and the event horizon $\mH^R$.}
\label{fig:RNdS}
\end{figure}
This spacetime describes a spherically symmetric charged BH in the presence of a positive cosmological constant. Its conformal diagram is depicted in Fig. \ref{fig:RNdS}. We will focus mostly on the region $\rI$, the exterior of the black hole, and the event horizon $\mH^R$. A consideration of region $\rII$, the inner horizon, and the relation to strong cosmic censorship can be found in \cite{PRLpaper}. 
The dynamics of the field $\Phi$ on this spacetime are described by the Klein-Gordon equation
\begin{subequations}
\begin{align}
\label{eq:KGE}
    \left[D_\nu D^\nu -\mu^2\right]\Phi&=0\\
    D_\nu&=\nabla_\nu-iqA_\nu\, .
\end{align}
\end{subequations}
Here, 
\begin{equation}
    A=-\frac{Q}{r}\text{d}t
\end{equation}
is the background electromagnetic field, and the covariant derivative $\nabla$ is taken with respect to the metric
\begin{subequations}
\begin{align}
    g&=-f(r) \text{d} t^2+f(r)^{-1} \text{d}r^2 + r^2 \text{d}\Omega^2\\
    f(r)&=-\frac{\Lambda}{3}r^2+1-\frac{2M}{r}+\frac{Q^2}{r^2} \end{align}
\end{subequations}
of our RNdS spacetime, where $M$, $Q$ and $\Lambda$ describe the mass and charge of the BH and the cosmological constant. The three positive real roots of $f$, $r_-<r_+<r_c$, mark the inner ($r_-$), event ($r_+$), and cosmological ($r_c$) horizon. 

To simplify \eqref{eq:KGE}, one can introduce the tortoise radial coordinate $r_*$, defined by $\text{d}r_*=f(r)^{-1}\text{d}r$. This coordinate approaches $-\infty$ near the event horizon and $+\infty$ near the inner and cosmological horizon. In addition to the change of variable, we employ the  mode ansatz
\begin{align}
\label{eq:ansatz}
    \Phi_{\ell m}(t,r_*,\theta,\phi)= (4\pi)^{-1/2} r^{-1}Y_{\ell m}(\theta ,  \phi )  h_{\ell}(r_*,t)\, 
\end{align}
for the scalar field. Here, the $Y_{\ell m}(\theta,\phi)$ are the spherical harmonics. If we assume 
\begin{align}
\label{eq:hSplit}
    h_{\ell}(r_*,t)=e^{-i\omega t}H_{\omega\ell}(r_*)\, ,
\end{align}
\eqref{eq:KGE} is reduced to an ordinary differential equation for $H_{\omega\ell}(r_*)$. It takes the form of a one-dimensional scattering problem
\begin{subequations}
\begin{align}
\label{eq:Diff_Eq}
    &\left[\partial_{r_*}^2+\left(\omega-\frac{qQ}{r}\right)^2-W\right]H_{\omega \ell}(r_*)=0\\
    &W=f(r)\left(\frac{\partial_rf(r)}{r}+\frac{\ell (\ell +1)}{r^2}+\mu^2\right)\, .
\end{align}
\end{subequations}
In terms of the tortoise coordinate $r_*$, the potential $W$ vanishes faster than any power as $r_*\to\pm\infty$.

In contrast to the real scalar field, the charged scalar field allows for gauge transformations of the form 
\begin{align}
\label{eq:gengauge}
    A_\nu &\to A_\nu+\partial_\nu \chi(x) &\Phi&\to e^{iq\chi(x)}\Phi \, .
\end{align}
We will fix our gauge partially and restrict to transformations of the form
\begin{equation}
\label{eq:red_gauge}
  \chi(t,r_*,\theta,\phi)=\tfrac{Q}{r_0}t\, .
\end{equation}
For $r_0=r_i$, $i\in \{+,-,c\}$, we denote the corresponding gauge by an $(i)$-superscript. The benefit of these gauges is that $A=0$ at the respective horizon, so that the functions $h_{\ell}(r_*,t)$ asymptotically behave as free waves for $r_*\to \pm\infty$ if the gauge is chosen accordingly. 

To extend the metric onto the horizons,  we introduce lightlike coordinates $v=t+r_*$, and $u=t-r_*$, and construct the Kruskal coordinates
\begin{align}
U&= \mp e^{-\kappa_+u}\,,  & V_c&= -e^{-\kappa_c v}\,, & V&= -e^{-\kappa_-v}\,,
\end{align}
with $\kappa_i=\tfrac{1}{2}\left|\partial_r f(r)\right|_{r=r_i}$, $i\in\{c,+,-\}$, the surface gravity of the corresponding horizon. In the definition of $U$, the upper sign is used  in $\rI$ and the lower sign is used in $\rII$. $V_c$ and $V$ are defined in $\rI$ and $\rII$, respectively.
%These coordinates can be smoothly extended across the horizons. 
Using $U$, we can smoothly extend the metric across $\mH^R$, and using $V$ ($V_c$) we can also extend it across $\mCH^R$ ($\mHc^L$). In these coordinates, the horizons $\mH^R$, $\mCH^R$, and $\mHc^L$ are situated at $U=0$, $V=0$, and $V_c=0$, respectively.

For the construction of the Unruh vacuum, we define a set of mode solutions $h^\ri_{\omega \ell}$, $h^\ru_{\omega \ell}$, called Unruh mode solutions, by initial conditions on $\mH^L \cup \mH^- \cup \mHc^- \cup \mHc^R$. 
%Initial conditions on $\mHc^R$ can be neglected, since events in $\rIII$ cannot impact the other regions. 
The Unruh mode solutions are of the form \eqref{eq:ansatz}, and the boundary conditions are
\begin{subequations}
\begin{align}
    h^{(c)\ri}_{\omega \ell}&=\begin{cases}
|\omega |^{-1/2} e^{-i\omega V_c} & \text{on }\mHc^- \cup \mHc^R \\ 0 & \text{on } \mH^-\cup \mH^L\\ \end{cases}\\
h^{(+)\ru}_{\omega \ell}&=\begin{cases}
0 &\text{on }\mHc^- \cup \mHc^R \\ |\omega |^{-1/2} e^{-i\omega U} & \text{on }\mH^-\cup\mH^L \, . \end{cases}
\end{align}
\end{subequations}

As mentioned in the Introduction, we assume the absence of classical instabilities or bound states in region $\rI$. This means that solutions of the form \eqref{eq:ansatz}, \eqref{eq:hSplit} with
\begin{subequations}
\label{eq:Instability}
\begin{align}
    h^{(+)}(r_*) & \sim e^{- i \omega v} &\text{ for } r_* \to - \infty, \\
    h^{(c)}(r_*) & \sim e^{- i (\omega + \ome) u} &\text{ for } r_* \to + \infty,
%    H^{(+)}_{\omega \ell}(r_*) & \sim e^{- i \omega r_*} & \text{ for } r_* \to - \infty, \\
%    H^{(c)}_{\omega \ell}(r_*) & \sim e^{ i \omega r_*} & \text{ for } r_* \to + \infty, 
\end{align}
\end{subequations}
where $\ome=qQ(r_+^{-1}-r_c^{-1})$, are only possible for frequencies $\omega$ with a negative imaginary part. In other words, the spectral gap of quasi-normal frequencies is strictly positive. As discussed above, this is the case except for a small parameter range of small $q, \mu, \Lambda$. If this condition is fulfilled,
%In the absence of bound states and instabilities, 
we can quantize the scalar field by an expansion in terms of the positive frequency Unruh modes as
\begin{align}
\Phi(x)&=\sum_{\lambda,\ell,m}\int\limits_{0}^{\infty}\text{d}\omega  \left(\Phi_{\omega \ell m}^{\lambda}(x) a_{\omega \ell m }^{\lambda} +\Phi_{-\omega \ell m}^{\lambda}(x)b_{\omega \ell m}^{\lambda \dagger}\right)\, ,
\end{align}
with $\lambda$ running over $\ri$ and $\ru$.  $a_{\omega\ell m}^{\lambda}$ and $b_{\omega\ell m}^\lambda$ are taken to be the usual Fock space annihilation operators, satisfying the canonical commutation relations. The Unruh vacuum  $|0\rangle_\rU$ is the ground state of this Fock space, for which $a^\lambda_{\omega\ell m}|0\rangle_\rU=0=b^\lambda_{\omega\ell m}|0\rangle_\rU$. As in \cite{Hollands:2019} and using results for decay of the field towards $i^-$ (c.f.\ Fig.~\ref{fig:RNdS}) \cite{Hintz:2015, Cardoso:2018}, one can show that this defines a proper quantum field/state (in the absence of classical instabilities or bound states).
%, provided that no classical instability is present (technically speaking, for a strictly positive spectral gap of quasinormal modes). As discussed in the Introduction, this is the case up to a small region in parameter space.
Furthermore, this state is Hadamard (see, for example, \cite{Hollands:2014} for a definition) in $\rI \cup \rII \cup \rIII$, i.e., across the cosmological and the event horizon. This is in contrast to the Boulware state considered in \cite{Balakumar:2020}, which can not be extended as a Hadamard state across the event horizon.

\section{The regularized current in the Unruh state}
\label{sec:current}

%In this section, we calculate an expression for the current of the charged scalar field in the Unruh vacuum. The classical version of that current
The observable that we focus on in this work is the current density $j_\nu$, which is classically given by
\begin{align}
\label{eq:current}
    j_\nu(x) = iq\left(\Phi(x)D_\nu^*\Phi^*(x)-\Phi^*(x)D_\nu \Phi(x)\right)\, 
\end{align}
with $D^*_\nu = \del_\nu + i q A_\nu$.
In the quantum theory, this needs to be renormalized, as it is quadratic and local in the field.  We do this using Hadamard point-split renormalization, which is local and (gauge) covariant \cite{Hollands:2001,Hollands:2014,Zahn:2012dz}. The expectation value of a renormalized field monomial $D_\alpha \Phi(x) D^*_\beta\Phi^*(x)$, with multi-indices $\alpha$ and $\beta$ in a Hadamard state $\Psi$, is defined as 
\begin{equation}
\label{eq:HadamardPointSplit}
 \langle D_\alpha \Phi(x) D^*_\beta\Phi^*(x)\rangle_\Psi^{\mathrm{ren}} = \lim\limits_{x^\prime\to x} D_\alpha (D^{\prime }_\beta)^{*}\left(\langle\Phi(x)\Phi^*(x^\prime)\rangle_\Psi - H(x,x^\prime)\right)\, .
\end{equation}
Here, $H(x,x^\prime)$ is the Hadamard parametrix of the wave operator $D_\nu D^\nu - \mu^2$, which is discussed in detail in the following section. For a Hadamard state $\Psi$, the expression in brackets on the right-hand side (r.h.s.)\ is smooth, so the limit of coinciding points is well defined and gauge invariant (before the limit of coinciding points is taken, the expression on the r.h.s.\ is of course not gauge invariant). 

Hadamard point-split renormalization is not the unique local and gauge covariant renormalization scheme. There are in general finite renormalization ambiguities. However, for the case of the current, any two locally gauge covariant renormalization schemes yielding a conserved current $j_\mu$ differ, at most, by multiple of the current $J_\mu$ responsible for the external electromagnetic field \cite{Zahn:2012dz,Balakumar:2019}, corresponding to a renormalization of the charge of the field. In the present case, the external current $J_\mu$ vanishes, so that Hadamard point-split renormalization yields a unique result. For some applications of this renormalization procedure in the context of external electromagnetic potentials or curved spacetimes, we refer to \cite{Schlemmer:2015sna,Wernersson:2020xeh} and \cite{Hollands:2014,Zilberman:2019,Hollands:2019}.

In this section, we focus on the evaluation of the first term on the r.h.s. of \eqref{eq:HadamardPointSplit}, in the linear combination of monomials relevant for the current density and evaluated in the Unruh state. Concretely, we first derive an expression for
\begin{equation}
\label{eq:Q_current}
    \langle j_\nu(x_1,x_2) \rangle_\rU =  iq \left\langle\left\{\Phi(x_1), \partial_\nu\Phi^{*}(x_2) \right\} - \left\{\Phi^{*}(x_1), \partial_\nu\Phi(x_2)\right\}\right\rangle_\rU .
\end{equation}
For convenience, we performed a symmetrization here, as $\{A,B\}=\tfrac{1}{2}(AB+BA)$.  In addition, we reduced the gauge-covariant derivative to a simple partial derivative by choosing a gauge of the form \eqref{eq:red_gauge}, with $r_0=r(x_2)$. 
%We will stick to this gauge for the rest of this section. 
As one can think of the point splitting as a regularization, we henceforth refer to this as the regularized current. We will further simplify it by choosing a particular prescription for the point splitting.

%It turns out that with a suitable 

%In order to do so, we will first compute a regularized expression for this current obtained via point-splitting. The regularized current can be written as
%\begin{align}
%\label{eq:Q_current}
%    \langle j_\nu(x_1,x_2) \rangle=&  iq\left\langle\left\{\Phi(x_1), \partial_\nu\Phi^{*}(x_2) \right\} - \left\{\Phi^{*}(x_1), \partial_\nu\Phi(x_2)\right\}\right\rangle\, .
%\end{align}
%Here, curly brackets denote anti-commutation, $\{A,B\}=\tfrac{1}{2}(AB+BA)$.  In addition, we reduced the gauge-covariant derivative to a simple partial derivative by choosing a gauge of the form \eqref{eq:red_gauge}, with $r_0=r(x_2)$. We will stick to this gauge for the rest of this section.

%After determining the  regularized current, we will compute its Hadamard parametrix. 
%Finally, taking the coincidence limit $x_1\to x_2$ in both parts, we find that both parts are finite separately. 

%Starting with the current, we introduce another type of mode solutions, called Boulware modes $\tilde{h}^{A}_{\omega\ell}$, for computational simplicity.
%$A$ runs over the two types of modes, $\ri$ and $\ru$. The modes are defined in region $\rI$ by initial conditions on $\mH^-\cup\mHc^-$, namely
As usual, it is convenient to re-express the Unruh modes $h^{\ri/\ru}_{\omega \ell}$ in terms of so-called Boulware modes $\tilde{h}^{\lambda \mathrm{N}}_{\omega\ell}$. Here $\lambda$ runs over $\{ \ri, \ru \}$, as before, while $\mathrm{N}$ runs over $\{ \rI, \rII \}$, labeling the region on which the modes are defined (as we will not be interested in evaluating the current density in region $\rIII$, we refrain from introducing Boulware modes for that region).
%$A$ runs over the two types of modes $\ri$ and $\ru$, combined with the regions $\rI$ and $\rII$ in which the modes are defined (as we will not be interested in evaluating the current density in region $\rIII$, we refrain from introducing Boulware modes for that region). Hence, the modes labelled $\ri\rI$ and $\ru\rI$ are defined on $\rI$ and the modes $\ri\rII$ and $\ru\rII$ on $\rII$. 
The boundary conditions for the modes in $\rI$ are defined on $\mH^-\cup\mHc^-$, and are given by
\begin{align}
\tilde{h}^{(c)\ri\rI}_{\omega\ell}&\sim e^{-i\omega v} \quad\text{on }\mHc^- \, , &  \tilde{h}^{(+)\ru\rI}_{\omega\ell}&\sim e^{-i\omega u} \quad\text{on } \mH^-\,,
\end{align}
with vanishing boundary conditions on the other horizon, respectively.
The modes in $\rII$ are defined on $\mH^L\cup\mH^R$, by the boundary conditions
\begin{align}
\tilde{h}^{(+)\ri\rII}_{\omega\ell}&\sim e^{-i\omega v}\quad \text{on }\mH^R \, , & \tilde{h}^{(+)\ru\rII}_{\omega\ell}&\sim e^{-i\omega u}\quad \text{on }\mH^L\,,
\end{align}
again with vanishing boundary conditions understood on the other part of the initial data surface. These boundary conditions are easier to use for computations than the ones for the Unruh modes. Note also that the modes defined in region $\rI$ can be continued to region $\rII$ by comparing their asymptotic behavior near $\mH^R$ with that of the modes defined in $\rII$.

%In order to express the Unruh modes in terms of the Boulware modes, we split the Unruh modes into the part defined by non-zero boundary conditions on $\mH^-\cup\mHc^-$ and the part which is non-trivial on $\mH^L$. 
%Here, we will only need the first part, and  we will call this part the $\ru$-type Unruh mode from now on.
%We can then express the Unruh modes as
In $\rI \cup \rII$, we can now express the Unruh modes in terms of the Boulware modes as
\begin{equation}
\label{eq:exp}
    |\omega |^{1/2} h^{\lambda}_{\omega\ell}(r_*,t)=  \sum\limits_\mathrm{N\in\{\rI,\rII\}}\int\limits_{-\infty}^\infty \frac{\text{d}\omega^\prime}{2\pi} \alpha^{\lambda\mathrm{N}}_{\omega\omega^\prime}\tilde{h}^{\lambda \mathrm{N}}_{\omega^\prime \ell}(r_*,t)\, ,
\end{equation}
where $\lambda \in\{\ri, \ru\}$. The coefficients $\alpha^{\lambda\mathrm{N}}_{\omega\omega^\prime}$ can be computed by applying the Fourier transform twice \cite{Lanir:2017}. 
%We can then express the Unruh modes as
%\begin{equation}
%\label{eq:exp}
%   h^{\lambda}_{\omega\ell}(r_*,t)=\int\limits_{-\infty}^\infty \frac{\text{d}\omega^\prime}{2\pi} \alpha^{\lambda}_{\omega\omega^\prime}\tilde{h}^{\lambda}_{\omega^\prime \ell}(r_*,t)\, ,
%\end{equation}
%where $\lambda \in\{\ri, \ru\}$ The coefficients $\alpha^{\lambda}_{\omega\omega^\prime}$ can be computed by applying Fourier transform twice \cite{Lanir:2017}. 
In the calculation, it is necessary to introduce a small imaginary part for $\omega^\prime$ as a regularization of a linear divergence for $\omega^\prime\to 0$. The regularized coefficients can be computed to be
\begin{subequations}
\begin{align}
\label{eq:alpha_w_w}
\alpha_{\omega \omega^\prime}^{\lambda\rI}&=\lim\limits_{\epsilon\to 0}\frac{1}{\kappa_j}|\omega|^{i\tfrac{\omega^\prime}{\kappa_j}}e^{\mathrm{sgn}(\omega)\tfrac{\pi\omega^\prime}{2\kappa_j}}\Gamma\left(-i\frac{\omega^\prime+i\epsilon}{\kappa_j}\right) \\
\alpha_{\omega\omega^\prime}^{\ru\rII}&=\overline{\alpha_{\omega\omega^\prime}^{\ru\rI}}\\
\alpha_{\omega\omega^\prime}^{\ri\rII}&=0
\end{align}
\end{subequations}
%\begin{align}
%\alpha_{\omega \omega^{'}}^{\lambda}&=\lim\limits_{\epsilon\to 0}\frac{1}{\kappa_j}|\omega|^{i\tfrac{\omega^{'}}{\kappa_j}}e^{\mathrm{sgn}(\omega)\tfrac{\pi\omega^{'}}{2\kappa_c}}\Gamma\left(-i\frac{\omega^{'}+i\epsilon}{\kappa_j}\right) 
%\end{align}
where $j=c$ for $\lambda=\ri$ and $j=+$ for $\lambda=\ru$. 

The expressions \eqref{eq:exp} for the Unruh mode functions can now be inserted into the expression for $\Phi(x)$. The regularized current \eqref{eq:Q_current} is then expressed in terms of the mode expansions of $\Phi(x)$, $\Phi^*(x)$, and their first partial derivatives. Since we are most interested in the $v$, $t$, and $r_*$-component of the current, we will assume that the partial derivative does not act on the $\theta$- and $\phi$- variables (the corresponding current components vanish by the rotational symmetry of the Unruh state). For $x_1=(r_1,t_1,\theta_1,\phi_1)$, and $x_2=(r_2,t_2,\theta_2,\phi_2)$, the expression for the regularized current \eqref{eq:Q_current} takes the form
\begin{align}
\label{eq:RegularizedCurrent}
    \langle j_\nu(x_1,x_2)\rangle_\rU & = \frac{q}{16\pi^3} \sum\limits_{\ell, m,\lambda,\mathrm{N},\mathrm{N}^\prime} \int\limits_0^\infty \frac{\text{d}\omega}{\omega} \int\limits_{-\infty}^{\infty}\text{d}\omega^\prime
    \int\limits_{-\infty}^{\infty}\text{d}\omega^{\prime\prime} \  \Imag \bigg[ \left( \overline{\alpha^{\lambda\mathrm{N}}_{\omega\omega^{\prime}}} \alpha^{\lambda\mathrm{N}^{\prime}}_{\omega\omega^{\prime\prime}} +\overline{\alpha^{\lambda\mathrm{N}}_{-\omega\omega^{\prime}}} \alpha^{\lambda\mathrm{N}^{\prime}}_{-\omega\omega^{\prime\prime}} \right)\times \\\nonumber
    &\times \frac{1}{r_1} \overline{Y_{\ell m}}(\theta_1,\phi_1)Y_{\ell m}(\theta_2,\phi_2)
   \overline{\tilde{h}_{\omega^\prime \ell}^{\lambda\mathrm{N}}}(r_1,t_1)\partial_\nu \left( \frac{1}{r_2}\tilde{h}^{\lambda\mathrm{N}^\prime}_{\omega^{\prime\prime}\ell}(r_2,t_2) \right) \bigg]\, ,
\end{align}
where the Boulware modes are evaluated in the gauge described below \eqref{eq:Q_current}.

Before continuing, we want to argue that the $i \epsilon$-prescription introduced in \eqref{eq:alpha_w_w}, which shifts the poles of $\omega^\prime$ and $\omega^{\prime\prime}$ in \eqref{eq:RegularizedCurrent}, can be safely dropped. This is due to the fact that the Boulware modes $\tilde{h}^\lambda_{\omega^\prime\ell}$ vanish as $\mathcal{O}(\omega^\prime)$ for $\omega^\prime\to 0$. To see this, let us first focus on the $\ru\rI$-modes. We find that they can be written in the form \eqref{eq:hSplit}.
The $\ru\rI$-modes enter region $\rI$ from $\mH^-$. From there, they can be scattered back to $\mH^R$, or travel all the way to $\mHc^L$. As a result, the asymptotic behavior of the radial part $H_{\omega\ell}(r_*)$ is
\begin{align}
\label{eq:asymp_H}
    H_{\omega\ell}\sim\begin{cases}
    e^{i\omega r_*}+\mathcal{R}_{\omega\ell}e^{-i\omega r_*} & r_*\to -\infty\\
    \mathcal{T}_{\omega\ell}e^{i(\omega+\ome)r_*} & r_*\to \infty\, ,
    \end{cases}
\end{align}
with $\ome$ as defined below \eqref{eq:Instability}.
%where $\ome=qQ(r_+^{-1}-r_c^{-1})$. 
There is no analytic expression for the scattering coefficients $\mathcal{R}_{\omega\ell}$ and $\mathcal{T}_{\omega\ell}$, but the behavior of $\mathcal{R}_{\omega\ell}$ for small $\omega$ can be estimated by a first-order approximation of the radial equation \eqref{eq:Diff_Eq} in $r-r_+$ near the event horizon \cite{Sela:2018}. One finds $\mathcal{R}_{\omega\ell}\sim -1+\mathcal{O}(\omega)$ for small $\omega$. Combining this result with the Wronskian relation
\begin{align}
\label{eq:TRrel}
    |\mathcal{R}_{\omega\ell}|^2+\frac{\omega+\ome}{\omega}|\mathcal{T}_{\omega\ell}|^2=1\, ,
\end{align}
of the scattering coefficients, one obtains that $\mathcal{T}_{\omega\ell}\sim \mathcal{O}(\omega)$ for small $\omega$. As a consequence, near the boundaries of region $\rI$, $H_{\omega\ell}$ vanishes point-wise as $\mathcal{O}(\omega)$ for $\omega\to 0$. The same is also true for $\partial_{r_*}H_{\omega\ell}(r_*)$. Since $H_{\omega\ell}$ satisfy \eqref{eq:Diff_Eq}, this implies $H_{\omega\ell}\to0$ point-wise for all $r_*$ as $\omega\to0$. A similar estimate can be found for the $\ri$-type modes on $\rI\cup\rII$, as well as for the combination of the reflected $\ru\rI$-modes and the $\ru\rII$-modes in $\rII$. As a result, the regularization of $\alpha_{\omega\omega^\prime}^{\lambda\mathrm{N}}$ 
can be dropped.

To proceed, we now compute the integral over $\omega$ in \eqref{eq:RegularizedCurrent}, which involves only the coefficients in the first line. One obtains results of the form\begin{align}
& \int\limits_0^\infty\frac{d\omega}{\omega} (\overline{\alpha_{-\omega\omega^\prime}^{\lambda\mathrm{N}}} \alpha_{-\omega\omega^{\prime\prime}}^{\lambda\mathrm{N}^{\prime}} +\overline{\alpha_{\omega\omega^\prime}^{\lambda\mathrm{N}}} \alpha_{\omega\omega^{\prime\prime}}^{\lambda\mathrm{N}^{\prime}})=4\pi^2C_{\lambda\mathrm{N}\mathrm{N}^\prime}(\omega^\prime)\delta(\omega^\prime-\omega^{\prime\prime})\, ,
\end{align}
where the $C_{\lambda\mathrm{N}\mathrm{N}^\prime}(\omega^\prime)$ are real functions given by
\begin{subequations}
\begin{align}
    C_{\ri,\rI,\rI}(\omega)& =\coth\left(\pi\tfrac{\omega}{\kappa_c}\right)\omega^{-1}\\
    C_{\ru,\rI,\rI}(\omega)= C_{\ru,\rII,\rII}(\omega) & = \coth\left(\pi\tfrac{\omega}{\kappa_+}\right)\omega^{-1}\\
    C_{\ru,\rI,\rII}(\omega)=C_{\ru,\rII,\rI}(\omega)& =\left[\omega\sinh\left(\pi\tfrac{\omega}{\kappa_+}\right)\right]^{-1}\, ,
\end{align}
\end{subequations}
and all other $C_{\lambda\mathrm{N}\mathrm{N}^\prime}(\omega)$ vanish.
Inserting this back in \eqref{eq:RegularizedCurrent}, we find
\begin{align}
\label{eq:RegularizedCurrent_2}
      \langle j_\nu(x_1,x_2)\rangle_\rU = &\frac{q}{4\pi r_1r_2} \sum\limits_{\ell, m,\lambda,\mathrm{N}\mathrm{N}^\prime} \int\limits_{-\infty}^{\infty} \text{d}\omega
      \ \Imag \bigg[ C_{\lambda\mathrm{N}\mathrm{N}^\prime}(\omega) \overline{Y_{\ell m}}(\theta_1,\phi_1) Y_{\ell m}(\theta_2,\phi_2)\times \\\nonumber
      & \times\overline{\tilde{h}_{\omega \ell}^{\lambda\mathrm{N}}}(r_1,t_1) \left(\partial_\nu\tilde{h}^{\lambda\mathrm{N}^\prime}_{\omega\ell}(r_2,t_2) -\frac{1}{r_2} \tilde{h}^{\lambda\mathrm{N}^\prime}_{\omega\ell}(r_2,t_2) \partial_\nu r_2\right)\bigg]\, .
\end{align}
%\begin{widetext}
%\begin{align}
%      \langle j_\nu(x_1,x_2)\rangle_U = &\frac{q}{4\pi r_1r_2} \sum\limits_{\ell, m,\lambda} \int\limits_{-\infty}^{\infty} \text{d}\omega
%      \text{Im}\left[C_{\lambda}(\omega) \overline{Y_{\ell m}}(\theta_1,\phi_1) Y_{\ell m}(\theta_2,\phi_2)\times\right.\\\nonumber
%      &\left.\times\overline{\tilde{h}_{\omega \ell}^{\lambda}}(r_1,t_1) \left(\partial_\nu\tilde{h}^{\lambda}_{\omega\ell}(r_2,t_2) -\frac{1}{r_2} \tilde{h}^{\lambda}_{\omega\ell}(r_2,t_2) \partial_\nu r_2\right)\right]\, .
%\end{align}
%\end{widetext}

So far, we have not specified the point splitting we apply. We use a $\theta$- splitting as introduced in \cite{Levi:2016}: We identify $r_2 = r_1$, $\phi_2 = \phi_1$, but set $\theta_2=\theta_1+\epsilon=\theta+\epsilon$ and in addition $t_2=t_1+\delta=t+\delta$ for the regularization of the $\omega$-integral. Note that we first take the limit $\delta\to 0$, before we take the limit $\epsilon \to 0$. This allows us to write
\begin{equation}
    \sum\limits_{m=-\ell}^{\ell}\overline{Y_{\ell m}}(\theta,\phi) Y_{\ell m}(\theta+\epsilon,\phi)=\frac{2\ell+1}{4\pi}P_\ell(\cos\epsilon)\, 
\end{equation}
for the $\omega$-independent part of the regularized current in \eqref{eq:RegularizedCurrent_2}.

Let us now take a closer look at the integral over $\omega$. Consider first the part with the derivative acting on $r$. In the coincidence limit, only the imaginary part of $\partial_\nu r$ will contribute to the integral. 
% ,due to the symmetry properties of the $C_{\lambda\mathrm{N}\mathrm{N}^\prime}$. 
However, $\partial_\nu r$ is real. Hence this part does not contribute to the integral in the coincidence limit, and we can drop it. The remaining integrals over $\omega$ are of the form
\begin{align}
\label{eq:w_Integral}
   \sum\limits_{\lambda,\mathrm{N},\mathrm{N}^\prime} \int\limits_{-\infty}^\infty \text{d}\omega \ C_{\lambda\mathrm{N}\mathrm{N}^\prime}(\omega) \Imag \left[ \overline{\tilde{h}_{\omega \ell}^{\lambda\mathrm{N}}}(r,t) \partial_\nu\tilde{h}^{\lambda\mathrm{N}^\prime}_{\omega\ell}(r,t+\delta)\right]\, .
\end{align}
%\begin{align}
%   \sum\limits_{\lambda} \int\limits_{-\infty}^\infty \text{d}\omega C_{\lambda}(\omega)\mathrm{Im}\left[ \overline{\tilde{h}_{\omega \ell}^{\lambda}}(r,t) \partial_a\tilde{h}^{\lambda}_{\omega\ell}(r,t+\delta)\right]\, .
%\end{align}
The splitting in $t$ will lead to an oscillatory term of the form $e^{-i\delta (\omega-\omega_g)}$, where $\omega_g$ is the frequency shift between the gauge in which the modes are defined and the one in which they are evaluated. To be able to take the limit $\delta \to 0$ later, we will shift the integrals by $\omega_g$.
This means that the integral involving the $\ri\rI$-type  Boulware mode will be shifted by $\omega_{r,\ri}=qQ(r^{-1}-r_c^{-1})$ and the one involving the $\ru\rI$- or $\ru\rII$-type mode will be shifted by $\omega_{r,\ru}=qQ(r^{-1}-r_+^{-1})$.

Next, consider the leading order in the large $|\omega|$-limit. There, \eqref{eq:Diff_Eq} becomes
\begin{align}
    \left[\partial_{r_*}^2+\omega^2\right]H_{\omega\ell}(r_*)=0\, 
\end{align}
so that the Boulware modes become plane waves. Then, the leading order contribution to the integrand of \eqref{eq:w_Integral}, evaluated in region $\rI$, is given by 
\begin{align}
    &\coth\left(\pi\tfrac{\omega}{\kappa_c}\right)  \left(\partial_\nu t+\partial_\nu r_*\right) \cos(\delta\omega)+\coth\left(\pi\tfrac{\omega}{\kappa_+}\right)  \left(\partial_\nu t-\partial_\nu r_*\right) \cos(\delta\omega)\, .
\end{align}
This integrand is antisymmetric in $\omega$, and the contributions from $\omega\to \infty$ and $\omega\to -\infty$ cancel. Similar results also hold for the integrand in $\rII$. The results for situations close to the present one \cite{Yafaev}, as well as the numerical results obtained with this formula, indicate that this cancellation is indeed strong enough to make the integral converge, even when the coincidence limit is taken. Assuming this to be true, we can take the limit $\delta\to0$, and hence 
\begin{align}
    \langle j_\nu(x)\rangle_\rU & = \lim\limits_{\epsilon\to0} \sum\limits_{\ell} \frac{q(2\ell+1)}{16\pi^2 r^2}P_\ell(\cos\epsilon)\int\limits_{0}^{\infty} \text{d}\omega  \sum\limits_{\lambda,\mathrm{N},\mathrm{N}^\prime} \left( C_{\lambda\mathrm{N}\mathrm{N}^\prime}(\omega+\omega_{r,\lambda}) \Imag \left[  \overline{\tilde{h}_{(\omega+\omega_{r,\lambda}) \ell}^{\lambda\mathrm{N}}}(r,t) \partial_\nu\tilde{h}^{\lambda\mathrm{N}^\prime}_{(\omega+\omega_{r,\lambda})\ell}(r,t) \right] + \omega \leftrightarrow - \omega \right) . 
\end{align}
%\begin{align}
%    \langle j_\nu(x)\rangle_\rU & = \lim\limits_{\epsilon\to0} \sum\limits_{\ell} \frac{q(2\ell+1)}{16\pi^2 r^2}P_\ell(\cos\epsilon)\int\limits_{-\infty}^{\infty} \text{d}\omega \sum\limits_{\lambda,\mathrm{N},\mathrm{N}^\prime} C_{\lambda\mathrm{N}\mathrm{N}^\prime}(\omega+\omega_{r,\lambda})\text{Im}\left[  \overline{\tilde{h}_{(\omega+\omega_{r,\lambda}) \ell}^{\lambda\mathrm{N}}}(r,t) \partial_\nu\tilde{h}^{\lambda\mathrm{N}^\prime}_{(\omega+\omega_{r,\lambda})\ell}(r,t) \right]\, . 
%\end{align}
%\begin{widetext}
%\begin{align}
%    \langle j_\nu(x)\rangle_U=& \lim\limits_{\epsilon\to0} \sum\limits_{\ell} \frac{q(2\ell+1)}{16\pi^2 r^2}P_\ell(\cos\epsilon)\int\limits_{-\infty}^{\infty} \text{d}\omega \sum\limits_{\lambda} C_{\lambda}(\omega+\omega_{r,\lambda})\text{Im}\left[ 
%   \overline{\tilde{h}_{(\omega+\omega_{r,\lambda}) \ell}^{\lambda}}(r,t) \partial_a\tilde{h}^{\lambda}_{(\omega+\omega_{r,\lambda})\ell}(r,t) \right]\, . 
%\end{align}
%\end{widetext}
From numerical computations, it turns out that this expression is already finite in the limit $\epsilon \to 0$. This is further confirmed by the calculation of the second term in \eqref{eq:HadamardPointSplit}, in the combination relevant for the current, which must also be, and indeed is, finite.

\section{The calculation of the parametrix}
\label{sec:ct}
%In this section, we want to compute the Hadamard parametrix for the current. 
%In the previous section, we calculated a regularized form of the current. In this section, we want to apply Hadamard renormalization. This is necessary even in the case when the parametrix is finite, since it implements the local and covariant normal ordering of the two fields $\Phi(x)$ in the current \cite{Hollands:2001}.

%The Hadamard renormalization works as follows: Consider the two-point function $\langle\nabla^\alpha \Phi(x)\nabla^\beta\Phi^*(x)\rangle_\Psi$, where $\Psi$ is a Hadamard state and $\alpha$ and $\beta$ are multi-indices. Then the renormalized form of this two-point function reads 
%\begin{align}
% &\langle\nabla^\alpha \Phi(x)\nabla^\beta\Phi^*(x)\rangle_\Psi^{\mathrm{ren}}\\\nonumber
% &=\lim\limits_{x^\prime\to x} \nabla^\alpha_x \nabla_{x^\prime}^\beta\left(\langle\Phi(x)\Phi^*(x^\prime)\rangle_\Psi-H(x,x^\prime)\right)\, .
%\end{align}
%Here, $H(x,x^\prime)$ is the Hadamard parametrix for the two-point function, which captures its state-independent singular behavior as $x^\prime\to x$. For a scalar field in four dimensions, it can be written as

We now turn to the discussion of the second term on the r.h.s.\ of \eqref{eq:HadamardPointSplit}. In four dimensions, the Hadamard parametrix can be written as
\begin{align}
 \label{eq:Hxxp}
    H(x,x^\prime)&= \frac{1}{8\pi^2} \left[ \frac{U(x,x^\prime)}{\sigma_{\epsilon}(x,x^\prime)} +\sum\limits_n V_n(x,x^\prime)\sigma^n(x,x^\prime) \ln\left(\frac{\sigma_{\epsilon}(x,x^\prime)}{K^2}\right)\right]\, .
\end{align}
Here $K$ is an arbitrary scale, and $\sigma_\epsilon(x,x')$ is Synge's world function \cite{Poisson:2011} equipped with an $i \epsilon$-prescription, which is only relevant for causally related $x$, $x'$ (and is thus irrelevant in the $\theta$-splitting that we are employing). The so-called Hadamard coefficients, $U(x,x')$ and $V_n(x,x')$, are smooth functions, which are determined in a local and (gauge) covariant way and fulfill
\begin{align}
\label{eq:HadamardSymmetry}
    \overline{U(x,x')} & = U(x',x), &
    \overline{V_n(x,x')} & = V_n(x',x).
\end{align}
The sum in \eqref{eq:Hxxp} does not converge in general, but for the application in the Hadamard point split \eqref{eq:HadamardPointSplit}, only a finite number of terms is relevant (whose number depends on the number of derivatives in the field monomial, i.e., $| \alpha |$ and $| \beta |$). From the requirement that the application of $D_\nu D^\nu - \mu^2$ yields a smooth function, one derives the transport equations
\begin{subequations}
\label{eq:TransportEquations}
\begin{align}
      \left[\sigma^\nu D_\nu + \frac{1}{2} \Box \sigma - 2 \right] U & = 0, \\
      2\left[\sigma^\nu D_\nu + \frac{1}{2} \Box \sigma - 1 \right]V_0 & = - \left[D_\nu D^\nu -\mu^2\right]U , \\
      2(n+1)\left[\sigma^\nu D_\nu + \frac{1}{2} \Box \sigma + n \right] V_{n+1} & = - [D_\nu D^\nu -\mu^2] V_n.
\end{align}
\end{subequations}
Here we used the notation $\sigma^\nu = \nabla^\nu \sigma$. That the leading short-distance singularity is as in the flat case is ensured by the initial condition $U(x,x) = 1$. This determines $U(x,x')$ uniquely as
\begin{equation}
     U(x,x^\prime)=\Delta^{\tfrac{1}{2}}(x,x^\prime)P(x,x^\prime)\, ,
\end{equation}
where $\Delta(x,x^\prime)$ is the Van Vleck-Morette determinant \cite{Poisson:2011}, and $P(x,x')$ is the parallel transport with respect to $D_\nu$ along the geodesic from $x'$ to $x$ (which can be assumed to be unique by restricting $x'$ to a suitably small neighborhood of $x$), i.e., solving $\sigma^\nu D_\nu P = 0$ with the initial condition $U(x,x) = 1$. The higher order Hadamard coefficients are then determined recursively.

For the application in Hadamard point splitting \eqref{eq:HadamardPointSplit}, only the coinciding point limit of a suitable number of derivatives of the Hadamard coefficients is relevant. More concretely, for the point-split renormalization of the current, we need to determine all the divergent and finite contributions to the coinciding point limit $x' \to x$ of $\partial'_\nu \Imag [H(x,x')]$, where we used \eqref{eq:HadamardSymmetry}. For this, we perform a Taylor expansion of the imaginary part of the Hadamard coefficients, in the form
%To solve these equations, we employ a covariant Taylor expansion of the form
\begin{align}
    F(x,x^\prime)&=F^{(0)}(x)+F^{(1)}_{\alpha}(x)\sigma^\alpha(x,x^\prime)+F^{(2)}_{\alpha\beta}(x)\sigma^\alpha(x,x^\prime)\sigma^\beta(x,x^\prime)+\dots
\end{align}
Evaluating successively higher covariant derivatives of $U$, $V_n$ in the limit of coinciding points, one obtains from the transport equations \eqref{eq:TransportEquations} the following results \cite{Balakumar:2019}:
\begin{subequations}
\begin{align}
    \Imag \left(U^{(1)}_\alpha\right)& = qA_\alpha\\
    \Imag \left(U^{(2)}_{\alpha\beta}\right) & = -\frac{q}{2}\nabla_{(\alpha}A_{\beta)}\\
    \Imag \left(U^{(3)}_{\alpha\beta\gamma}\right) & = \frac{q}{6} \Real \left(D_{(\alpha}D_\beta A_{\gamma)}\right) + \frac{q}{12}A_{(\alpha}R_{\beta\gamma)}\\
    \Imag \left(V_{0\alpha}^{(1)}\right) & = \frac{q}{2}\left[\mu^2-\frac{1}{6}R\right]A_\alpha-\frac{q}{12} \nabla^\nu F_{\nu\alpha}\\
   \Imag \left(V_{0\alpha\beta}^{(2)}\right) & = - \frac{q}{4} \left[\mu^2-\frac{1}{6}R\right] \Real \left(D_{(\alpha}A_{\beta)}\right)+\frac{q}{24}A_{(\alpha}\nabla_{\beta)}R -\frac{q}{24}\nabla_{(\alpha}\nabla^{\nu}F_{\beta)\nu} \\
   \Imag \left(V^{(0)}_1\right)&=0\, ,
\end{align}
\end{subequations}
where symmetrization with respect to the indices in round brackets is understood.
In addition, we will need an expansion of $\sigma^\mu$ in terms of the angular separation $\epsilon$. For this, we use the expansion given in \cite{Ottewill:2008}, and find
\begin{subequations}
\label{eq:SigmaExp}
\begin{align}
    \sigma&=\frac{r^2}{2}\epsilon^2- \frac{fr^2}{24}\epsilon^4+\mathcal{O}\left(\epsilon^6\right)\\
    \sigma_\theta&=-r^2\epsilon+\frac{fr^2}{6}\epsilon^3+\mathcal{O}\left(\epsilon^4\right)\\
    \sigma_r&=\frac{r}{2}\epsilon^2+\mathcal{O}\left(\epsilon^4\right)\\
    \sigma_t&\approx\sigma_\varphi=0+\mathcal{O}\left(\epsilon^4\right)\, .
\end{align}
\end{subequations}

Taking into account that we are in a gauge such that $A(x)=A(x^\prime)=0$, we can now put these ingredients together and compute the coinciding point limit of $\partial'_\nu \Imag[ H(x,x')]$ in the $\theta$-splitting, which is the relevant quantity for our purpose of renormalizing the current density. We find that the divergent parts cancel, consistent with our finding in the previous section, and for the finite part we obtain
\begin{equation}
\label{eq:parametrix}
    -2q \partial'_{\nu} \Imag \left[H(x,x^\prime)\right]= -\frac{1}{4\pi^2} \frac{q^2Qf(r)}{6r^3} \delta^t_\nu +\mathcal{O}(\epsilon)\, .
\end{equation}
We note that this term is only relevant for the $t$-component of the current and vanishes on the horizons. Similar finite contributions from Hadamard point splitting frequently occur in the renormalization of the current density (see the discussion and examples in \cite{Schlemmer:2015sna,Wernersson:2020xeh}).
%Note that this term is finite in the coincidence limit, and that it vanishes as one approaches any of the horizons. 
Comparing this expression to the result of \cite{Herman:1995}, where a divergence in the coinciding point limit is found, we note that there is an error in the calculation of \cite{Herman:1995}: in their formula for the current, Eq. (8), the difference rather than the sum of contributions with derivatives acting on the first and second variable
%of $G^{(1)}$ 
should be taken. If this is taken into account, the results do agree. 

Note also that the parametrix satisfies the condition for the conservation of the current, as can be shown by an explicit calculation in the $(t,r_*,\theta,\phi)$ coordinates. Employing that by spherical symmetry and stationarity the angular components of the current must vanish, $j_\theta=j_\phi=0$, and that it must be time-independent, the current conservation $\nabla_\nu j^\nu=0$ reduces to $\partial_{r_*}(r^2j_{r_*})=0$ \cite{Balakumar:2020}. Since the parametrix only affects the $t$-component of the current and is time-independent, the current is conserved if the expression without the parametrix yields a conserved current.

Finally, our expression for the renormalized  current is given by subtracting the Hadamard parametrix from the result of the previous section. Since this parametrix is finite, and supported by the numerical results, we take the limit $\epsilon\to 0$. We find
%\begin{widetext}
\begin{align}
\label{eq:J_ext}
   \langle j_\nu(x)\rangle_\rU & = \sum\limits_{\ell} \frac{q(2\ell+1)}{16\pi^2 r^2}\int\limits_{0}^{\infty} \text{d}\omega \sum\limits_{\lambda,\mathrm{N},\mathrm{N}^\prime} \left( C_{\lambda\mathrm{N}\mathrm{N}^\prime}(\omega+\omega_{r,\lambda}) \Imag \left[ 
   \overline{\tilde{h}_{(\omega+\omega_{r,\lambda}) \ell}^{\lambda\mathrm{N}}}(r,t) \partial_\nu\tilde{h}^{\lambda\mathrm{N}^\prime}_{(\omega+\omega_{r,\lambda})\ell}(r,t) \right] + \omega \leftrightarrow - \omega \right) \nonumber \\
   & \quad +\frac{1}{4\pi^2}\frac{q^2Qf(r)}{6r^3}\delta^t_\nu .
\end{align}
%\end{widetext}
%\begin{widetext}
%\begin{align}
%    \langle j_\nu(x)\rangle_U=&  \sum\limits_{\ell} \frac{q(2\ell+1)}{16\pi^2 r^2}\int\limits_{-\infty}^{\infty} \text{d}\omega \sum\limits_{\lambda} C_{\lambda}(\omega+\omega_{r,\lambda})\text{Im}\left[ 
%   \overline{\tilde{h}_{(\omega+\omega_{r,\lambda}) \ell}^{\lambda}}(r,t) \partial_a\tilde{h}^{\lambda}_{(\omega+\omega_{r,\lambda})\ell}(r,t) \right]+\frac{1}{4\pi^2}\frac{q^2Qf}{6r^3}\delta^t_\nu
%\end{align}
%\end{widetext}

\section{Numerical results}
\label{sec:num}
While there is no analytic solution for the functions $\tilde{h}^{\lambda N}_{\omega\ell}(r_*,t)$, they can be approximated numerically. This allows us, for example, to calculate the $v$-component of the current on the event horizon. Here, we can use the asymptotic behavior of the Boulware modes \eqref{eq:asymp_H}
to get an expression in terms of scattering coefficients. We find
\begin{subequations}
\begin{align}
\label{eq:J_EH}
    \langle j_v\rangle_\rU &= \sum\limits_{\ell} \frac{q(2\ell+1)}{16\pi^2 r^2}\int\limits_{0}^{\infty} \text{d}\omega \ \left( F(\omega) + F(-\omega) \right) \\
    F(\omega)&=\coth\left(\pi\tfrac{\omega+\ome}{\kappa_c}\right)\left(1-|\mathcal{R}_{\omega\ell}|^2\right)+\coth\left(\pi\tfrac{\omega}{\kappa_+}\right)|\mathcal{R}_{\omega\ell}|^2  \, .
\end{align}
\end{subequations}
Due to the fact that the radial part of the Klein-Gordon equation can be brought into the form of a Heun equation in the conformally coupled case, $\mu^2=2\Lambda/3$, \cite{Suzuki:1999, Hollands:2020}, this case can be dealt with most efficiently. However, more general masses can be handled as well \cite{Hollands:2020}: one can obtain solutions to \eqref{eq:Diff_Eq} which are defined in the neighborhood of one of the three horizons. Comparing these solutions on the overlap of their domains, one can then compute the scattering coefficients. In this way, the scattering coefficients and the integrand in \eqref{eq:J_EH} or \eqref{eq:J_ext} are computed for different values of $\omega$ and $\ell$ up to some maximal values, beyond which the contributions become negligibly small. The integral is finally estimated using a Riemann sum. The sum is also used to estimate the error of this procedure. For more details, see \cite{Hollands:2020}.

Note that in order to achieve a good performance of the code, we restrict ourselves to small charges $qQ$ and masses $\mu^2$ of the scalar field. As discussed above, we thus need to choose the cosmological constant $\Lambda$ large enough in order to avoid the instability parameter region. Equation (46) in \cite{Dias:2018} gives a condition to determine whether the instability appears for a chosen set of spacetime parameters. We checked this condition and find no instability in the parameter range we consider. 
%We have explicitly checked the formula for the leading imaginary part of the frequency of the $\ell=\mu^2=\omega_0=0$-mode, with $\omega_0$ the frequency at $qQ=0$, for small $qQ$, eq. (46) in \cite{Dias:2018}. We find no instability in the parameter range we consider. 

\begin{figure}[h]
    \centering
    \includegraphics[width=0.6\textwidth]{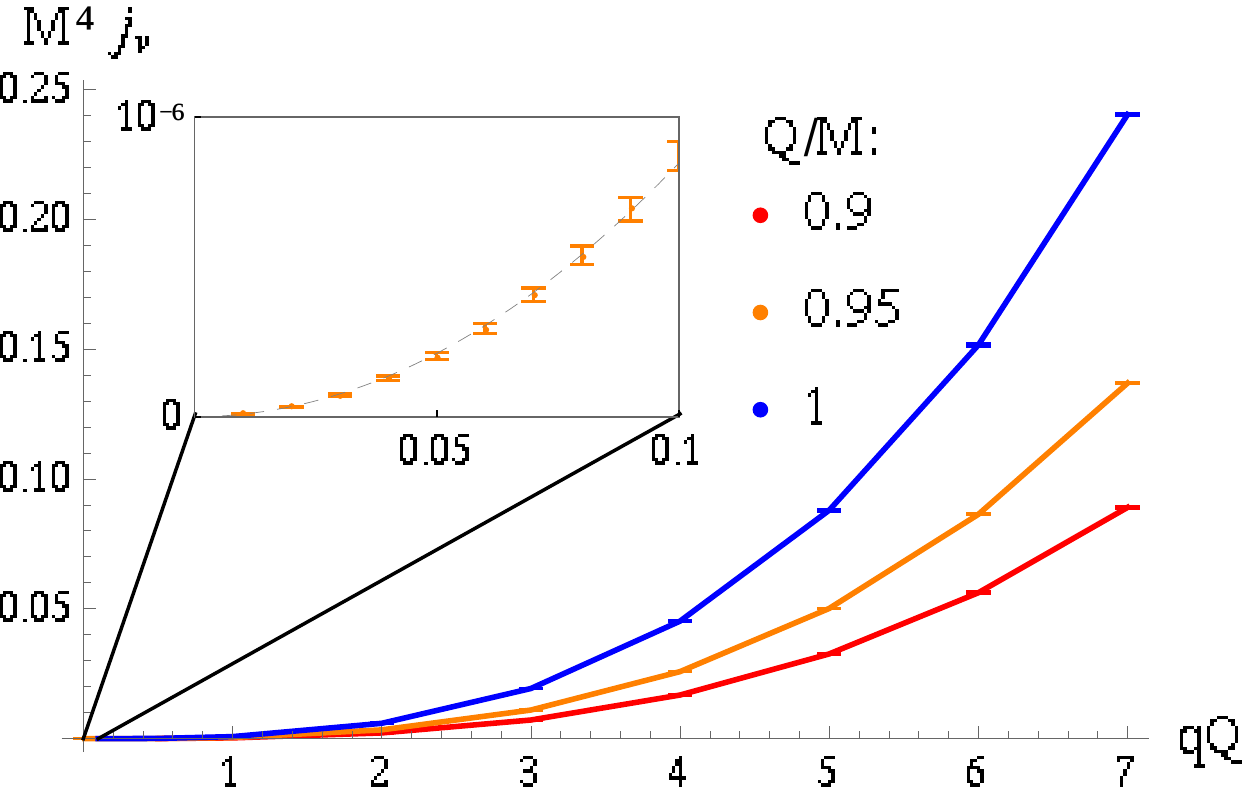}
    \caption{The $v$-component of the current of a conformally coupled scalar field on the event horizon as a function of the scalar field charge in the Unruh vacuum for $Q=0.95 M$ and $\Lambda=0.14 M^{-2}$. The smaller graph shows the results for $Q=0.95M$ for small $qQ$, the dashed line indicates a $q^2$-fit.}
    \label{fig:q}
\end{figure}
\begin{figure}[h]
    \centering
    \includegraphics[width=0.6\textwidth]{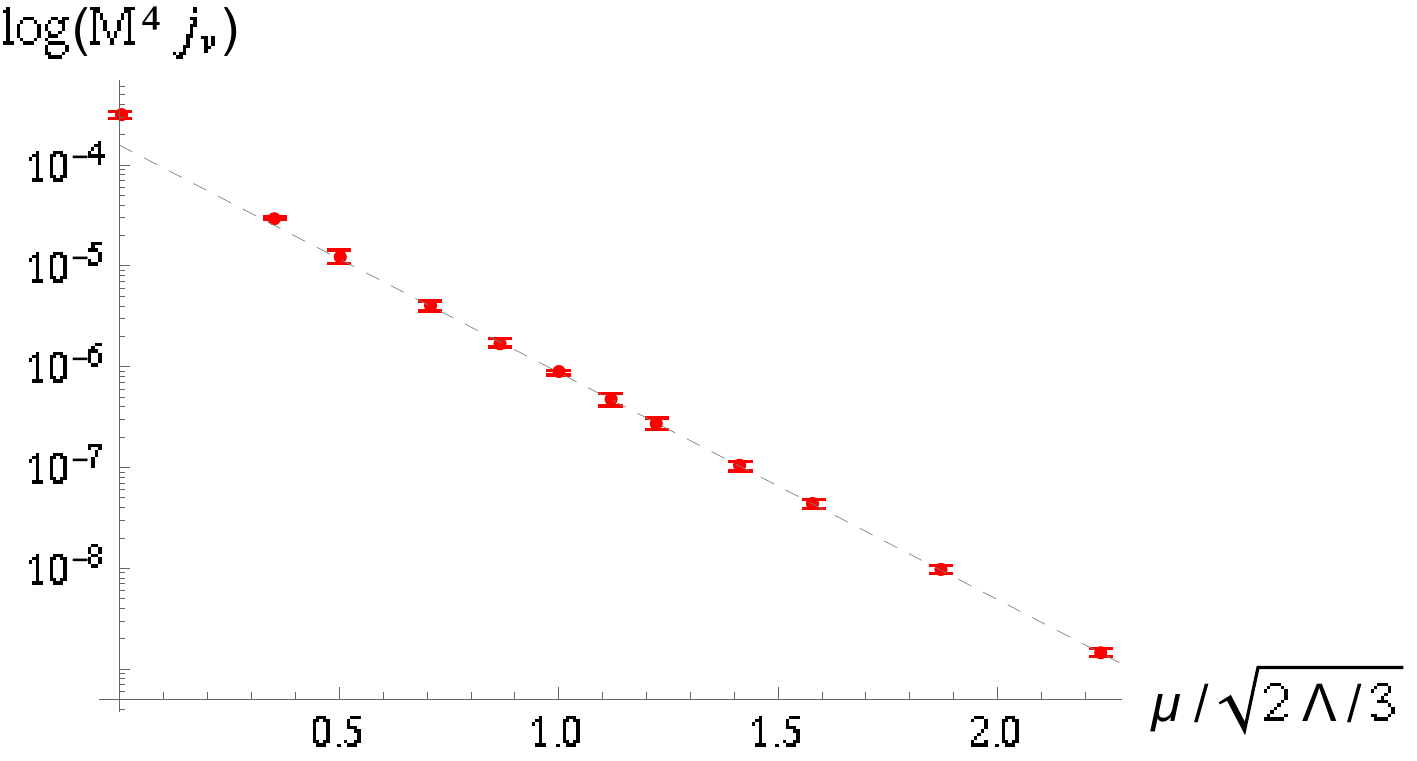}
    \caption{The $v$-component of the current of a charged scalar field of charge $qQ=0.1$ at the event horizon as a function of the scalar field mass in the Unruh vacuum for $Q=0.95 M$ and $\Lambda=0.14 M^{-2}$.}
    \label{fig:m}
\end{figure}
Figures \ref{fig:q} and \ref{fig:m} show the $v$-component of the  current on the event horizon. 
In both plots, the cosmological constant is chosen as $\Lambda=0.14 M^{-2}$, for comparability with \cite{PRLpaper}. 
In Fig. \ref{fig:q} the current is plotted as a function of the scalar charge $qQ$ for a conformally coupled scalar field and for different values of $Q/M$. We observe that $\langle j_v\rangle_\rU$ increases with $Q/M$. This is in line with the results of \cite{Chen:2012}, who found that the pair-production rate of charged scalars by an extremal BH is larger than that of a near-extremal one. In addition, we find that for small charges $q$ of the scalar field the current behaves as $q^2$. This can be seen in the smaller graph in Fig. \ref{fig:q}, where the results for $Q=0.95$ and $qQ\leq0.1$ have been fitted with $a\times(qQ)^2$, shown by the dashed line. This is seemingly in contraction with Schwinger's pair creation formula \cite{Schwinger:1951} widely used in the literature on black hole discharge, \cite{Zaumen:1974,Carter:1974,Damour:1974}, according to which the pair creation rate $\Gamma$ should be nonperturbative in $q$ near $q = 0$, involving a factor 
\begin{equation}
\label{eq:Schwinger}
    \Gamma \sim \exp\left[-\frac{\pi\mu^2r_+^2}{qQ}\right]\, .
\end{equation}
For the parameters used for Fig.~\ref{fig:q}, $\pi \mu^2 r_+^2 \sim 0.84$. Thus, the effect of such an exponential suppression should be clearly visible in the embedded figure. That this is not the case indicates that \eqref{eq:Schwinger} is not applicable for small (conformal) mass $\mu$ and small charge $q$. This is not very surprising, as for conformal mass the Compton length coincides with the Hubble length. Intuitively, the proportionality $\sim q^2$ can be understood as follows: a charge $q$ created near the horizon is accelerated away from the BH with acceleration proportional to $q$. Its contribution to the current is thus $\sim q^2$.

In Fig.~\ref{fig:m}, the current is displayed as a function of the mass $\mu$ in units of the conformal mass $\sqrt{2\Lambda/3}$, for $qQ=0.1$ and $Q=0.95M$. We find that $\langle j_v\rangle_\rU$ decays exponentially with $\mu$, and the corresponding fit is represented by the dashed line (the data point of the massless case is excluded from the fit). This also seems to be in contradiction to \eqref{eq:Schwinger}, which would yield an exponential decay in $\mu^2$ rather than $\mu$. This matches our expectation that the Schwinger formula is not applicable for masses of the order of $\sqrt{2\Lambda/3}$.
%Let us compare the results found here to the Schwinger formula \cite{Schwinger:1951} widely used in the literature on black hole discharge, \cite{Zaumen:1974,Carter:1974,Damour:1974}. The Schwinger formula giving the number of pairs created per unit volume applied to our setup can be written as \cite{Schwinger:1951,Herman:1994}
%\begin{equation}
%    \Gamma=\frac{q^2Q^2}{4\pi^3r_+^4}\exp\left[-\frac{\pi\mu^2r_+^2}{qQ}\right]\, .
%\end{equation}
%We fitted the functional dependence of the Schwinger formula on $q$ and $\mu^2$ to our results for the current.
%We find that the functional dependence on $\mu$ of the Schwinger formula can describe our results quite well, at least for masses $\gtrsim\Lambda$. For $\mu^2\gtrsim \Lambda$, the current decreases approximately exponentially with $\mu^2$, as one would expect from the Schwinger formula \eqref{eq:Schwinger}. However, the rate of decay is roughly one order of magnitude smaller than expected from \eqref{eq:Schwinger}. Furthermore, and not surprisingly, for very small masses $\mu^2$ deviations from the exponential law can be observed. 
%Thus, our results indicate that applying the Schwinger formula near the event horizon of a black hole is justified as long as $\mu^2$ is sufficiently large. This also shows very nicely that our results for the current near the horizon behave as expected.

As the Unruh state is Hadamard across $\mH^R$, the renormalized expectation value $\langle j_U \rangle_\rU$ must be finite on $\mH^R$. By the tensor transformation law, it follows that $\langle j_u \rangle_\rU$ vanishes on $\mH^R$. This implies that on the event horizon $\langle j_v\rangle_\rU = \langle j_{r_*}\rangle_\rU = \langle j_t\rangle_\rU$. Hence the results presented above also apply to $\langle j_t\rangle_\rU$. In particular, we find that near the horizon the charge density is negative and its magnitude decreases exponentially as the Compton-wavelength $1/\mu$ decreases.

%Note that on $\mH^-$, and hence by stationarity of our state also on $\mH^R$, we have $\langle j_u\rangle_U=0$. This implies that on the horizon $\langle j_v\rangle_U=\langle j_{r_*}\rangle_U=\langle j_t\rangle_U$. Hence the results presented above also apply to the charge density $\langle j_t\rangle_U$. In particular, we find that the charge density decreases exponentially as the Compton-wavelength $1/\mu$ decreases.

\begin{figure}
    \centering
    \includegraphics[width=0.6\textwidth]{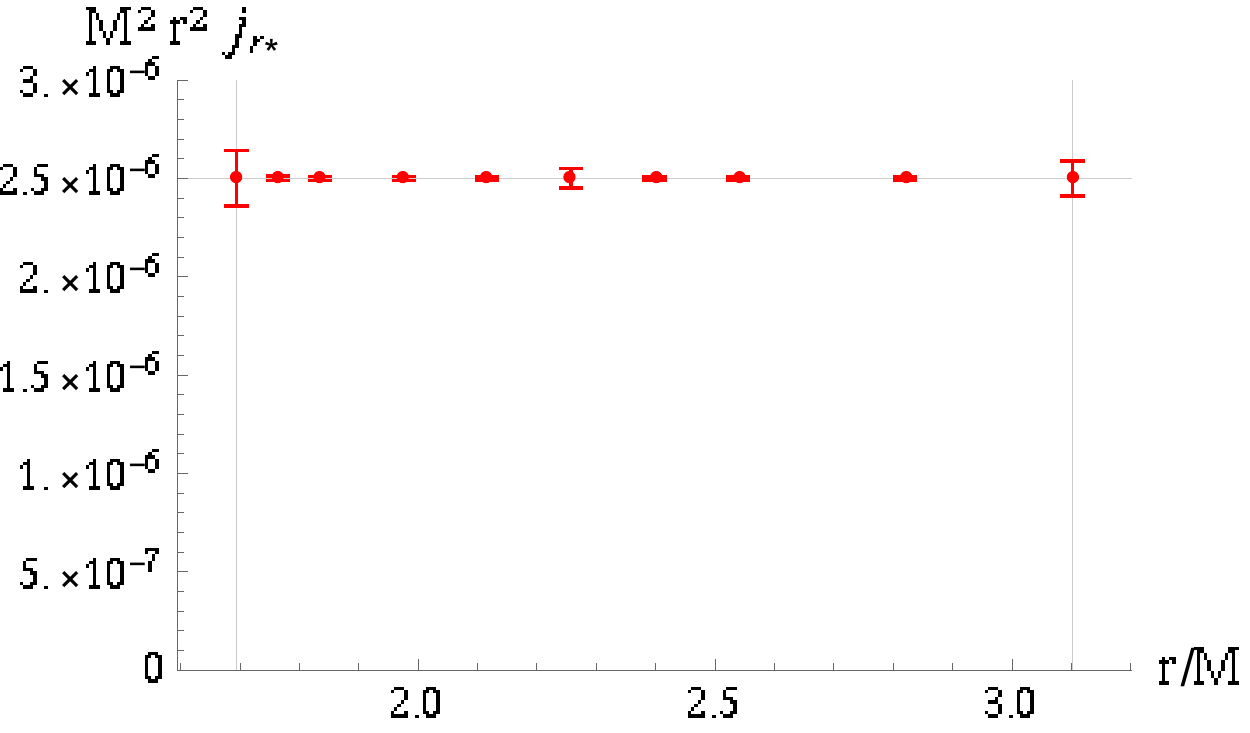}
    \caption{$r^2\langle j_{r_*}\rangle_\rU$ for the conformally coupled scalar at different radii $r$. This quantity must be constant for the current conservation. The vertical lines mark $r_+$ and $r_c$, respectively, where the first and last point are computed. $Q=0.95 M$, $\Lambda=0.14 M^{-2}$, and $qQ=0.1$.}
    \label{fig:cons}
\end{figure}

With our general formula \eqref{eq:J_ext}, we can also calculate the current in the interior of region $\rI$. This can for example be used to check the conservation of the current by calculating $r^2\langle j_{r_*}\rangle_\rU$ at different values of the radius $r$. If the current is conserved, the result must be a constant that agrees with $r_+^2\langle j_v\rangle_\rU|_{\mH}$. The results for $qQ=0.1$, $\Lambda=0.14M^{-2}$ and $Q=0.95M$ are shown in Fig. \ref{fig:cons}. We find that our results satisfy the condition for current conservation very well.

Furthermore, we can compute $\langle j_t\rangle_\rU$, related to the charge density, in the exterior region. This result is the hardest to obtain numerically. In particular, estimating the errors of our calculation via the Riemann sum is insufficient in this case,  since it strongly overestimates the inaccuracy. Instead, we carefully consider all error sources separately. First of all, we calculate the integral in \eqref{eq:J_ext} only up to some $\omega_{\mathrm{max}}$. The rest of the integral can, however, be approximated by fitting the exponential decay we expect for large $\omega$ to the last few $\omega\leq\omega_{\mathrm{max}}$, and adding the integral of the fitted function from $\omega_{\mathrm{max}}$ to $\infty$ to our result. For our chosen parameters, the correction term is smaller than $3\times10^{-11}$ for all $\ell$ considered. Second, we approximate the integral by a Riemann sum. To estimate the error made by this discretization, we compare the result of the Riemann sum with the full number of points to the result obtained by using only every second point. We find that the change in the result is less than $10^{-10}$ for all values of $\ell$ considered, becoming slightly larger as $\ell$ increases. Finally, we only consider $\ell$ up to some finite $\ell_{\mathrm{max}}$. To estimate the error due to this cutoff, we observe that for all $\ell$ taken into account, the contribution to the current decreases by a factor $k<0.9$ as $\ell$ increases, and the contribution from $\ell_{\mathrm{max}}$ is of the order $10^{-10}$. Due to this analysis, we estimate that the uncertainty of our results is about $10^{-8}-10^{-9}$, which corresponds to $1-10\%$. 

\begin{figure}
    \centering
    \includegraphics[width=0.6\textwidth]{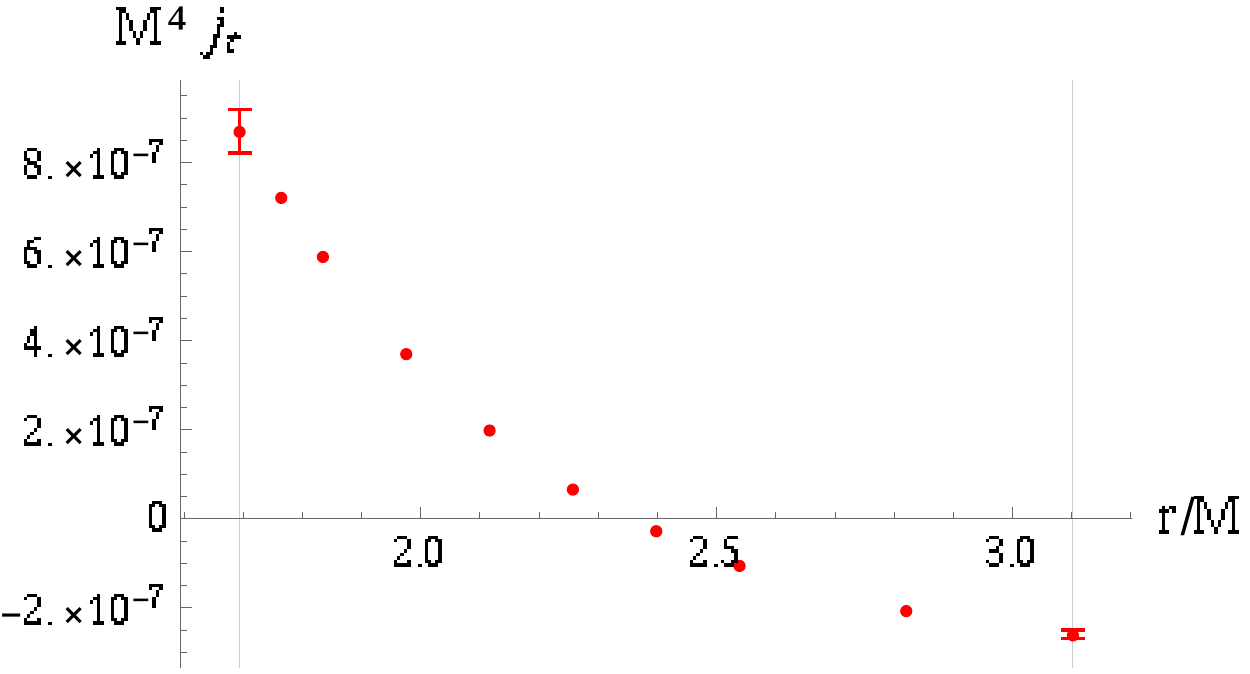}
    \caption{$\langle j_t\rangle_\rU$ for the conformally coupled scalar in region $\rI$ as a function of $r$ for $qQ=0.1$, $Q=0.95 M$, and $\Lambda=0.14 M^{-2}$. The first and last point are results obtained at $r_+$ and $r_c$ respectively, including realistic errors. $r_+$ and $r_c$ are also marked by the vertical lines.}
    \label{fig:jt}
\end{figure}

The results are shown in Fig. \ref{fig:jt}. Note that the first and last point shown in Fig.~\ref{fig:jt} are results obtained on $\mH^R$ and $\mHc^L$ respectively, including conservative errors estimated as detailed in \cite{Hollands:2020}. For the results in the interior, we omitted the error bars. However the data points seem to interpolate reasonably between the two horizons.
We observe that $\langle j_t\rangle_\rU$ starts positive near the event horizon and becomes negative as one approaches the cosmological horizon. 
This is to be expected in the Unruh state, which is Hadamard across the cosmological horizon: Using the same argument as for the $u$ component on $\mH^R$, we find that $\langle j_v \rangle_\rU$ vanishes on $\mHc^L$.
%This is expected from our construction of the Unruh vacuum: We can calculate the $v$-component of the current on $\mHc^-$, and find that it vanishes there by our choice of state. 
Hence, we must have $\langle j_t\rangle_\rU=-\langle j_{r_*}\rangle_\rU$ on the cosmological horizon. However, by current conservation and our previous results, $\langle j_{r_*}\rangle_\rU=Kr_c^{-2}$ on $\mHc$, where $K=\langle r^2j_{r_*}\rangle_\rU$ is a positive constant. Hence $\langle j_t\rangle_\rU<0$ on the cosmological horizon. This is also to be expected on physical grounds: in region $\rIII$, $\del_t$ is spacelike and directed inwards, so that $-j_t$ is the outward current. For a positively charged BH, we indeed expect this to be positive.

\section{Conclusion}
\label{sec:fin}
We derived a formula for the current of a charged scalar field in the RNdS spacetime, using point splitting and Hadamard renormalization. We found that the Hadamard parametrix only contributes a finite term, in contrast to results obtained previously \cite{Herman:1995}. In addition, this term vanishes on all horizons. We obtained an explicit expression for the current of the charged scalar in the RNdS spacetime, which can also be used in the interior of the BH, as was done in \cite{PRLpaper}. We tested the result by evaluating it numerically. We evaluated the $v$-component of the current on $\mH^R$ in the Unruh state and found an exponential decay in the mass $\mu$, as well as a quadratic dependence on the charge $q$ of the field for small $q$. We also calculated the $r_*$-component of the current in the exterior region of the spacetime to test current conservation, and found it to hold. Finally, $\langle j_t\rangle_\rU$, related to vacuum polarization, was computed inside region $\rI$, with results compatible with those obtained on the horizons bounding region $\rI$.  
%The resulting expression for the current can now be used also to study the interior of the black hole.

\section*{Acknowledgements}
C.K. wants to thank Stefan Hollands for suggesting this topic. This work has been funded by the Deutsche Forschungsgemeinschaft (DFG) under the Grant No. 406116891 within the Research Training Group RTG 2522/1.

\bibliography{cs}
\end{document}